# The $N_2O$-$CS_2$ dimer is cross-shaped

N. Moazzen-Ahmadi, [1]

A.R.W. McKellar, [2]

[1] *Department of Physics and Astronomy, University of Calgary, 2500 University Drive North West, Calgary, Alberta T2N 1N4, Canada.*

[2] *National Research Council of Canada, Ottawa, Ontario K1A 0R6, Canada.*

**Abstract**

The infrared spectrum of the cross-shaped van der Waals complex $N_2O - CS_2$ is observed in the region of the $N_2O$ $\nu_1$ fundamental band ($\approx 2220$ cm$^{-1}$) using a tuneable diode laser to probe a pulsed supersonic slit jet expansion. Both $^{14}$N- and $^{15}$N-substituted species are studied. Analysis of their spectra establishes that this dimer has a cross-shaped structure, similar to its isoelectronic cousin $CO_2 - CS_2$. This is the first spectroscopic observation of $N_2O$-$CS_2$, and the molecular parameters determined here should be useful for detection of its pure rotational microwave spectrum.

**1. Introduction**

It has been found that most weakly-bound dimers containing small linear polyatomic monomers have planar structures [1], following from the prototypical case of the $CO_2$ dimer [2,3]. Thus it was a bit of a surprise when, in 1998, Dutton et al. [4] showed that the $CO_2 - CS_2$ dimer was non-planar and cross-shaped, as viewed along the intermolecular axis. More recently, similar cross-shaped structures have been reported as well for the $CS_2$ dimer [5] and for one isomer of $OCS - CS_2$ [6]. In the present paper, we show that $N_2O - CS_2$ is also cross-shaped (see Fig. 1) by analysing its infrared spectrum in the region of the $N_2O$ $\nu_1$ fundamental band.

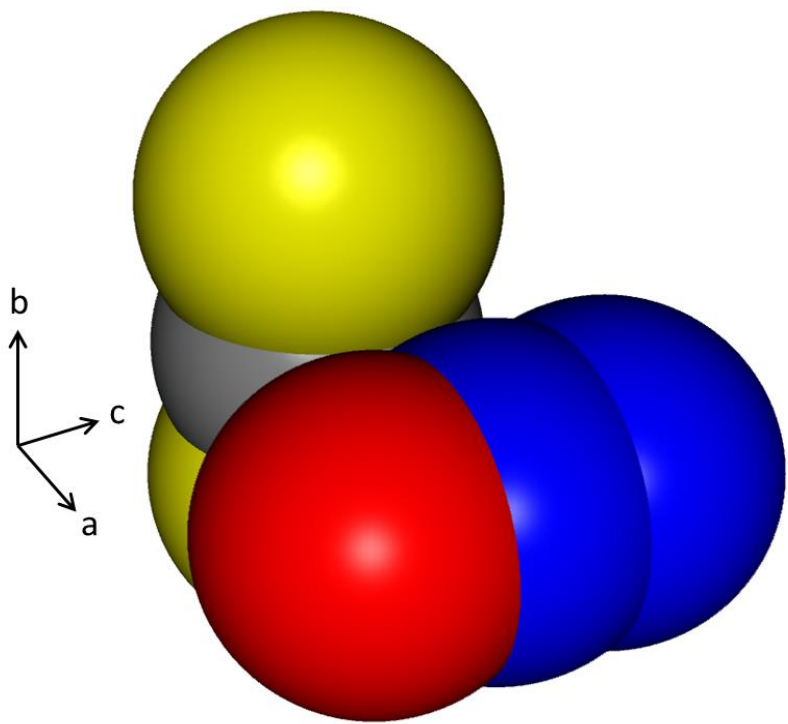

Fig. 1. Illustration of the cross-shaped structure of $N_2O - CS_2$, showing the direction of the inertial axes.

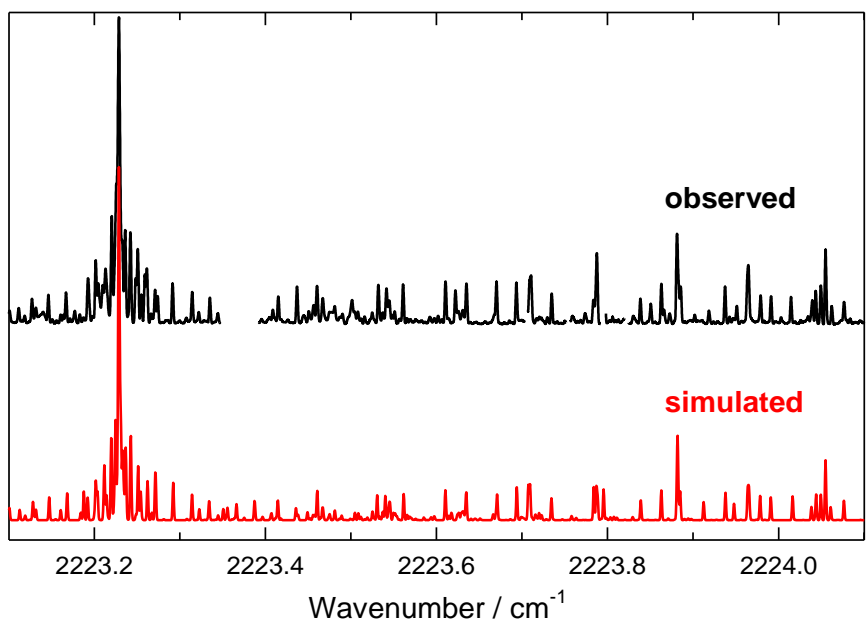

Fig. 2. Observed and simulated spectra of $^{14}N_2O - CS_2$ in the region of the $N_2O$ $\nu_1$ fundamental. The simulation uses an effective rotational temperature of 2.5 K and a Gaussian linewidth of 0.0016 cm$^{-1}$. Blank sections of the observed spectrum were obscured by $N_2O$ monomer absorption.

**2. Observed spectra**

A pulsed supersonic slit jet apparatus was used with a lead salt tuneable diode laser as probe [1]. The expansion gas was a very dilute mixture of about 0.15% $N_2O$ plus 0.3% $CS_2$ in helium. The backing pressure of about 10 atmospheres resulted in an effective rotational temperature around 2.5 K. Wavenumber calibration was made using a fixed etalon and a room temperature $N_2O$ reference gas cell. The PGOPHER computer package was used for spectral simulation and fitting [7].

The $N_2O$ and $CO_2$ molecules are rather similar in terms of mass and intermolecular forces. We thus expected $N_2O$ - $CS_2$ to have a similar cross-shaped structure, similar rotational parameters, and a similar *c*-type band as compared to $CO_2$ – $CS_2$ [4,8]. This was indeed the case, but there is an important difference between the two dimers: $CO_2 - CS_2$ has a $C_2$ symmetry axis which results in half the rotational levels being missing, while $N_2O - CS_2$ has no such symmetry.

Part of the observed $N_2O - CS_2$ spectrum is shown in Fig. 2. A strong central *Q*-branch at 2223.23 cm$^{-1}$ is flanked by weaker *P*- and *R*-branches. The band is similar to that of $CO_2 - CS_2$ (see Fig. 2 of [8]), but denser and more symmetric due to the lack of missing levels. We had relatively little trouble assigning the spectrum since the rotational parameters were indeed similar to those of $CO_2 - CS_2$. However, there were some discrepancies in the analysis which we attributed to upper state perturbations. For this reason, we decided to analyse the ground state first using combination differences from the spectrum, so that it would not be contaminated by any upper state perturbations. A total of 63 ground state combination differences with values of *J* and $K_a$ up to 9 and 6, respectively, were thus fitted with an average (root mean square) error of about 0.0003 cm$^{-1}$. Then the ground state parameters were held fixed while those of the upper state were fitted to the assigned infrared line positions. A total of 134 lines were fitted in terms of 167 transitions (some lines are blended) with an average error of about 0.0006 cm$^{-1}$. The parameters resulting from these two fits are listed in Table 1. It should be noted that the 1σ uncertainties given here for the upper state are probably unrealistically small since the lower state parameters were fixed in the infrared fit.

We were also able to observe the analogous band for $^{15}N_2O - CS_2$ using an enriched sample of $^{15}N_2O$, and the central part of this spectrum is illustrated in Fig. 3. Details of the analysis and fit were very similar to those of the normal isotopologue, and results are given in Table 1. The ground state fit involved 75 combination differences and had an average error of about 0.0003 cm$^{-1}$. The infrared fit involved 135 lines, 184 transitions, and an average error of about 0.0009 cm$^{-1}$.

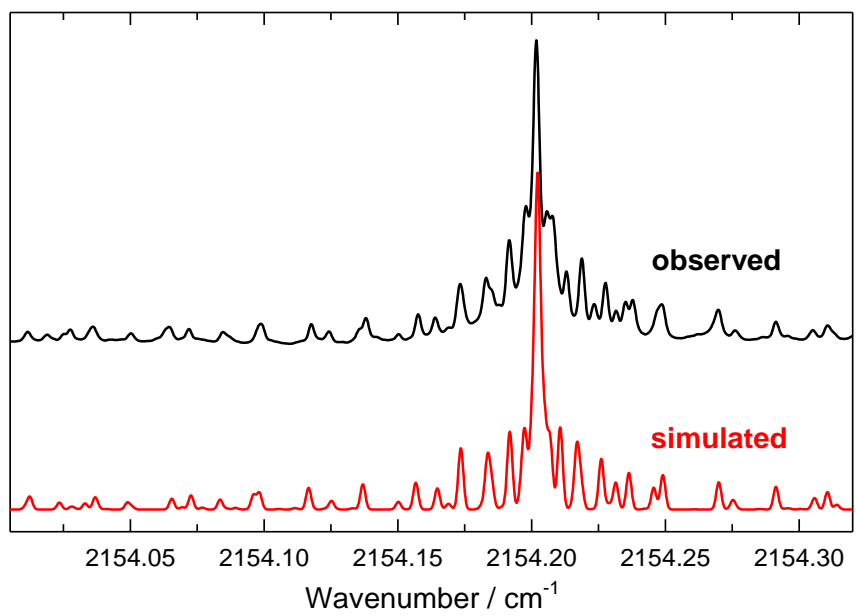

Fig. 3. Observed and simulated spectra of $^{15}N_2O - CS_2$ in the region of the $N_2O$ $\nu_1$ fundamental.

Table 1. Parameters for the ground state and excited states of $N_2O$-$CS_2$, compared with those of $CO_2 - CS_2$ [8] (values in cm$^{-1}$).[a]

|  | $^{14}N_2O - CS_2$ | $^{15}N_2O - CS_2$ | $CO_2 - CS_2$ |
|---|---|---|---|
| $\nu_0$ | 2223.2320(1) | 2154.2053(2) | 2346.5463(1) |
| $A'$ | 0.086997(16) | 0.086406(11) | 0.085763(13) |
| $B'$ | 0.047330(7) | 0.046021(6) | 0.046342(15) |
| $C'$ | 0.035726(6) | 0.035082(6) | 0.035405(20) |
| $D_K'$ | -9.0(37) e-7 | -17.1(35) e-7 [b] | -8.6(26) e-7 |
| $D_{JK}'$ | -0.7(26) e-7 | 4.6(20) e-7 | 8.0(31) e-7 |
| $D_J'$ | 4.11(44) e-7 | 5.54(30) e-7 | 2.87(98) e-7 |
| $A''$ | 0.087081(17) | 0.086536(11) | 0.085916(15) |
| $B''$ | 0.047407(13) | 0.046095(13) | 0.046380(14) |
| $C''$ | 0.035764(14) | 0.035119(14) | 0.035403(21) |
| $D_K''$ | -5.1(57) e-7 | -4.4(37) e-7 | -8.9(26) e-7 |
| $D_{JK}''$ | -21.7(60) e-7 | -6.4(42) e-7 | 7.8(31) e-7 |
| $D_J''$ | 5.7(17) e-7 | 7.1(16) e-7 | 3.30(91) e-7 |

[a] Quantities in parentheses are statistical errors (1σ) in units of the last quoted digit. GSCDs were fit for the ground state, hence uncertainties for the excited state are unrealistically small (see text).

Table 2. Observed vibrational shifts of $N_2O - CS_2$ and $CO_2 - CS_2$ (values in cm$^{-1}$).[a]

|  | Shift |
|---|---|
| $^{14}N_2O - CS_2$<br>$N_2O$ $\nu_1$ band | -0.525 |
| $^{15}N_2O - CS_2$<br>$N_2O$ $\nu_1$ band | -0.520 |
| $CO_2 - CS_2$<br>$CO_2$ $\nu_3$ band | -2.597 |

[a] The $\nu_1$ mode of $N_2O$ (2223.76 cm$^{-1}$ for $^{14}N_2O$) and $\nu_3$ mode of $CO_2$ (2349.14 cm$^{-1}$) are roughly analogous.

## 3. Discussion and conclusions

As shown in Table 2, the observed band origins of $N_2O - CS_2$ are shifted by only a relatively small amount ($\approx$-0.5 cm$^{-1}$) from those of the free $N_2O$ monomer, showing that the presence of the nearby $CS_2$ molecule has little effect on the frequency of the $N_2O$ $\nu_1$ mode. Interestingly, a considerably larger red shift of -2.6 cm$^{-1}$ is observed for $CO_2 - CS_2$ [4,8] in the somewhat analogous $CO_2$ $\nu_3$ mode.

We can also compare the rotational parameters of $N_2O - CS_2$ and $CO_2 - CS_2$, as shown in Table 1. Looking at $^{14}N_2O - CS_2$, we find that the rotational constants $A$, $B$, and $C$ are increased by 1.36, 2.21, and 1.02 %, respectively, relative to $^{12}C^{16}O_2 - CS_2$ (which has the same reduced mass). The increases are partly due to the fact that the $B$-value of $N_2O$ itself is larger than that of $CO_2$ by 7.4%. But, in addition, it turns out that the intermolecular bond is slightly shorter for $N_2O - CS_2$. Determining this bond length as described in [4], and averaging the slightly different values obtained using $B''$ or $C''$, we obtain values of 3.368 and 3.367 Å for $^{14}N_2O - CS_2$ and $^{15}N_2O - CS_2$, respectively, as compared to 3.392 Å for $CO_2 - CS_2$ [4].

But there is an additional structural parameter to determine in the case of $N_2O$ – $CS_2$, namely the angle between the intermolecular axis (connecting the monomer centers of mass) and the $N_2O$ (or $CO_2$) axis. For $CO_2$ – $CS_2$, we know by symmetry that this angle has an equilibrium value of 90º, but for $N_2O$ – $CS_2$ this is not necessarily the case. If $CO_2$ – $CS_2$ had a rigid cross-shaped structure, then its *A*-value (0.08592 cm$^{-1}$) should be the inverse sum of the $CO_2$ and $CS_2$ *B*-values (0.08526 cm$^{-1}$). The fact that *A* is slightly (0.00066 cm$^{-1}$) larger is a reflection of the (zero-point) intermolecular bending motions of $CS_2$ and $CO_2$ within the dimer. That is, the angles between the intermolecular and monomer axes are not always 90º, even though these are the average values. Turning to $N_2O$ – $CS_2$, if the equilibrium angle between the intermolecular and $N_2O$ axes were significantly different from 90º, then we might expect this to show up as a larger than expected *A*-value, assuming similar intermolecular bending amplitudes in the two dimers. However, it turns out that the difference between *A* and the inverse sum of the *B*s is quite similar, namely 0.00052 cm$^{-1}$ for $^{14}N_2O$ – $CS_2$ and 0.00059 cm$^{-1}$ for $^{15}N_2O$ – $CS_2$. So there is no strong evidence for this angle being significantly different from 90º in $N_2O$ – $CS_2$. A similar conclusion is reached from the absence of any observed *a*-type transitions in the bands observed here (deviation from 90º would project some of the $N_2O$ stretching transition moment onto the dimer *a*-axis).

It should be possible in the future to observe the pure rotational microwave spectrum of $N_2O$ – $CS_2$ and thus obtain more precise rotational parameters and possibly improved structural data. We would expect *c*-type transitions arising from the small permanent dipole moment of $N_2O$, as well as possible *a*-type transitions arising from a small induced dipole. The expected transition frequencies can be easily calculated from our ground state parameters in Table 1. One problem is that the intensity of each transition is likely to be spread out among many components due to resolved hyperfine splitting in $^{14}N_2O - CS_2$. In this respect, $^{15}N_2O - CS_2$ would be more promising, if an enriched sample is available.

In conclusion, the weakly bound van der Waals complex $N_2O - CS_2$ has been observed spectroscopically for the first time by means of its infrared spectrum in the region of the $N_2O$ $\nu_1$ stretch. Both $^{14}N_2O$- and $^{15}N_2O$-containing dimers were studied, and the structure was shown to be cross-shaped and very similar to that of the isoelectronic species $CO_2$-$CS_2$. The ground state parameters determined here should be useful for future observation of the microwave spectrum of $N_2O - CS_2$.

**Acknowledgements**

We gratefully acknowledge the financial support of the Natural Sciences and Engineering Research Council of Canada.

**References**

[1] N. Moazzen-Ahmadi, A.R.W. McKellar, Int. Rev. Phys. Chem. 32 (2013) 611.

[2] K.W. Jucks, Z.S. Huang, D. Dayton, R.E. Miller, W.J. Lafferty, J. Chem. Phys. 86 (1987) 4341.

[3] M.A. Walsh, T.H. England, T.R. Dyke, B.J. Howard, Chem. Phys. Lett. 142 (1987) 265.

[4] C.C.Dutton, D.A. Dows, R. Eikey, S. Evans, R.A. Beaudet, J. Phys. Chem. A 102 (1998) 6904.

[5] M. Rezaei, J. Norooz Oliaee, N. Moazzen-Ahmadi, A.R.W. McKellar, J. Chem. Phys. 134 (2011) 144306.

[6] J. Norooz Oliaee, F. Mivehvar, M. Dehghany, N. Moazzen-Ahmadi, J. Phys. Chem. A 114 (2010) 7311.

[7] PGOPHER version 8.0, C. M. Western, 2014, University of Bristol Research Data Repository, doi:10.5523/bris.huflggvpcuc1zvliqed497r2

[8] M. Dehghany, M. Rezaei, N. Moazzen-Ahmadi, A.R.W. McKellar, J. Brown, X.-G. Wang, T. Carrington, Jr., J. Mol. Spectrosc. 330 (2016) 188.